\documentstyle[prl,aps]{revtex}

\def\err#1#2{\stackrel{\scriptstyle +#1}{\scriptstyle -#2}}
\begin{document}
\draft

\title{Constraints on Electron-quark Contact Interactions and Implications to
models of leptoquarks and Extra $Z$ Bosons}
\author{Kingman Cheung
\footnote{Email: cheung@phys.cts.nthu.edu.tw}
}
\address{National Center for Theoretical Science, National Tsing Hua 
University, Hsinchu, Taiwan, R.O.C.}
\date{\today}
\maketitle

\begin{abstract}
We update the global constraint on four-fermion $ee
q q$ contact interactions.
In this update, we included the published data of H1 and ZEUS for the 
94--96 run in the $e^+ p$ mode and the newly published data of H1 for
the 1999 run in the $e^- p$ mode. 
Other major changes are the new LEPII data on hadronic cross sections 
above 189 GeV, and the atomic parity violation measurement on Cesium 
because of a new and improved atomic calculation, 
which drives the data within $1\sigma$ of the standard model value.
The global data do not show any evidence for contact interactions, and 
we obtain 95\% C.L. limits on the compositeness scale.
A limit of $\Lambda^{eu}_{LL+(-)} > 23\, (12.5)$ TeV is obtained.  
Implications to models of leptoquarks and extra $Z$ bosons are examined.
\end{abstract}
\pacs{\hfill{NSC-NCTS-010622}}

\section{Introduction}

Four-fermion contact interaction is not something new, but was proposed 
decades ago by Fermi to account for the nuclear beta decay. 
The interaction is represented by
\[
{\cal L} \sim G_F \left( \bar e \gamma^\mu (1-\gamma^5) \nu \right)\;
                  \left( \bar u \gamma_\mu (1-\gamma^5) d \right) \;
\]
where $G_F$ is the Fermi constant with dimension $[{\rm mass}]^{-2}$.
This interaction is not renormalizable because the amplitude grows
indefinitely with the energy scale if $G_F$ is kept constant.  
It was only until
60's that the electroweak theory was proposed.  The four-fermion contact 
interaction was then replaced by an exchange of the weak gauge boson $W$ and
$G_F$ replaced by the $W$ boson propagator: $G_F \to 1/(p^2 - m_W^2)$.
The weak gauge bosons were only
discovered later when the energy scale reached the hundred GeV level.
In the above history we learn a couple of lessons:
(i) the existence of four-fermion contact interactions is a signal of
new physics beyond the existing standard theory, and 
(ii) the exact nature of new physics is unknown at the low energy 
scale.  Only when the energy scale is high enough can the nature of new 
physics be probed.

Previous and present collider experiments have been searching for signs of
four-fermion contact interactions, including experiments at the Tevatron,
at HERA, at LEP, and at low-energy $e$-N and $\nu$-N scattering experiments
(for a summary see Refs. \cite{ours,others}.)  
If deviations from the standard model (SM) were seen
this would be a clear indication of new physics and would drive our 
resources towards where the new physics belongs.

A global analysis of the neutral-current (NC) $ee q q$ data was
performed three years ago \cite{ours}, 
\footnote{A model-independent analysis was previously done in Ref. \cite{sch}
for studying new physics at HERA.}
which was motivated by the
HERA anomaly \cite{H1-97,zeus-97}, in which H1 and ZEUS recorded a significant
excess in NC deep-inelastic cross sections in the high-$Q^2$ region.
The advantage of analyzing the $ee qq$ contact interactions is that
they can show up in a number of channels, therefore 
we can use global NC data sets to put bounds on $ee qq$
contact interactions.

Since then the data collected by H1 and ZEUS 
in 1997 agreed well the SM.  We performed 
updates in 1998 \cite{update-1998} based on their preliminary data.
Since our previous fits, there have been some other changes in the data sets,
which will likely affect the fit, and so we update the analysis in this note.
The changes are summarized as follows.
In 1999, both H1 and ZEUS published their final data \cite{zeus,H1} on NC 
deep-inelastic scattering in the $e^+ p$ mode, which show
good agreement with the SM, except that ZEUS still has two high $Q^2$ events
where  only 0.2 is expected.  H1 also published their new data in the 
1999 run in $e^- p$ mode \cite{H1}.
Other  major changes are the new LEPII data on hadronic cross sections 
at energies above 189 GeV \cite{lepew}, and 
the atomic parity violation (APV) measurement on Cesium 
\cite{apv} because of a new and improved atomic calculation \cite{atom}, 
which drives the data within $1\sigma$ of the SM value \cite{langacker}.
Note that the hadronic cross sections given by the LEP Electroweak Working 
Group \cite{lepew} showed a $2.5\sigma$ deviation above the SM predictions.

The purpose of this note is to update the analysis 
that examines the NC data sets from current accelerator experiments to see 
if there is any sign of contact interactions.
If so it is a signal of new physics; if not we put limits
on the compositeness scale $\Lambda$.

In the next section, we describe the formalism and followed by the descriptions
of various data sets in Sec. III. We present the fits and limits in Sec. IV.  
In Sec. V and VI, we  extend the analysis to models of
leptoquarks and extra $Z$ bosons, respectively.  We conclude in Sec. VII.

\section{Parametrization}

The conventional effective Lagrangian of $e e q q$ contact
interactions has the form \cite{eich}
\begin{eqnarray}
L_{NC} &=& \sum_q \Bigl[ \eta^{eq}_{LL}
\left(\overline{e_L} \gamma_\mu e_L\right)
\left(\overline{q_L} \gamma^\mu q_L \right)
+ \eta^{eq}_{RR} \left(\overline{e_R}\gamma_\mu e_R\right)
                 \left(\overline{q_R}\gamma^\mu q_R\right) \nonumber\\
&& \quad + \eta^{eq}_{LR} \left(\overline{e_L} \gamma_\mu e_L\right)
                             \left(\overline{q_R}\gamma^\mu q_R\right)
+ \eta^{eq}_{RL} \left(\overline{e_R} \gamma_\mu e_R\right)
\left(\overline{q_L} \gamma^\mu q_L \right) \Bigr] \,, \label{effL}
\end{eqnarray}
where eight independent coefficients $\eta_{\alpha\beta}^{eu}$ and
$\eta_{\alpha\beta}^{ed}$ have dimension (TeV)$^{-2}$ and are conventionally
expressed as $\eta_{\alpha\beta}^{eq} = \epsilon g^2 /\Lambda_{eq}^2$,
with a fixed $g^2=4\pi$.
The sign factor $\epsilon= \pm 1$ allows for either constructive or destructive
interference with the SM $\gamma$ and $Z$ exchange amplitudes and
$\Lambda_{eq}$ represents the mass scale of the exchanged new
particles, with coupling strength $g^2/4\pi=1$. A coupling of this
order is expected in substructure models and $\Lambda_{eq}$ is
often called the ``compositeness scale''.

In models with SU(2)$_L$ symmetry, we expect some relations among the 
contact interaction coefficients.  The particle content has
the left-handed leptons and quarks in SU(2) doublets $L=(\nu_L,e_L)$
and $Q=(u_L,d_L)$, while the right-handed electrons and quarks in singlets.
The most general $SU(2)_L \times U(1)$ invariant contact term
Lagrangian is given by
\begin{eqnarray}
{\cal L}_{SU(2)}&=&
\eta_1 \Bigl(\overline L\gamma^\mu L\Bigr) \Bigl(\overline Q\gamma_\mu Q\Bigr)
+
\eta_2 \Bigl(\overline L\gamma^\mu T^aL\Bigr) \Bigl(\overline Q\gamma_\mu T^a
Q\Bigr) \nonumber \\
&& + 
\eta_3 \Bigl(\overline L\gamma^\mu L\Bigr)
\Bigl(\overline{u_R}\gamma_\mu
u_R\Bigr)
+ \eta_4 \Bigl(\overline L\gamma^\mu L\Bigr)
\Bigl(\overline{d_R}\gamma_\mu
d_R\Bigr) \nonumber \\
&&+ 
\eta_5 \Bigl(\overline{e_R}\gamma^\mu e_R\Bigr) \Bigl(\overline Q\gamma_\mu
Q\Bigr) +
\eta_6 \Bigl(\overline{e_R}\gamma^\mu e_R\Bigr) \Bigl(\overline{u_R}\gamma_\mu
u_R\Bigr) \nonumber\\
&& + \eta_7 \Bigl(\overline{e_R}\gamma^\mu e_R\Bigr)
\Bigl(\overline{d_R}\gamma_\mu
d_R\Bigr)\; .
\label{effLsu2}
\end{eqnarray}
By expanding the $\eta_5$ term we have 
\begin{equation}
\eta^{eu}_{RL}=\eta_5=\eta^{ed}_{RL} \; .
\label{su2releR}
\end{equation}
In addition, the four neutrino and the lepton couplings are also 
related by SU(2):
\begin{equation}
\eta^{\nu u}_{LL}  = \eta^{ed}_{LL}\; , 
\eta^{\nu d}_{LL}  = \eta^{eu}_{LL}\; , 
\eta^{\nu u}_{LR}  = \eta^{eu}_{LR}\; , 
\eta^{\nu d}_{LR}  = \eta^{ed}_{LR}\; . 
\label{su2relnu}
\end{equation}
In our analysis, the relations of Eqs.~(\ref{su2releR}) and (\ref{su2relnu})
are only used when neutrino scattering data are included in the analysis. 
We shall state clearly when these SU(2) relations are used or not.  This 
is because in some combinations of $\eta$'s, at least one of the SU(2) 
relations cannot be held, then we are forced not to use the SU(2) symmetry.

Even though we expect that $SU(2)_L \times U(1)$ will be a
symmetry of the renormalizable interactions which ultimately manifest
themselves
as the contact terms of Eq.~(\ref{effL}), electroweak symmetry breaking may
break the mass degeneracy of SU(2) multiplets of the 
heavy quanta that 
give rise to (\ref{effL}). This would result in a violation of the relations
of Eqs.~(\ref{su2releR}) and (\ref{su2relnu}). 

Because of severe
experimental constraints on intergenerational transitions like
$K\to\mu e$
we restrict our discussions to first generation contact terms. Only
where required by particular data (e.g. the muon sample of Drell-yan
production at the Tevatron) shall  we assume universality of contact terms 
between $e$ and $\mu$.

Let us start with the scattering process $q\bar q \to \ell^+ \ell^-$ ($\ell=
e,\mu$).
The amplitude squared for $q\bar q \to \ell^+ \ell^-$ or
$\ell^+ \ell^- \to q\bar q $ (without averaging initial spins or colors) is
given by
\begin{equation}
\sum |{\cal M}|^2 = 4 u^2 \left( |M^{\ell q}_{LL}(s)|^2 + 
     |M^{\ell q}_{RR}(s)|^2 \right )
  + 4 t^2 \left( |M^{\ell q}_{LR}(s)|^2 + 
|M^{\ell q}_{RL}(s)|^2 \right ) \;,
\end{equation}
where
\begin{equation}
\label{mab}
M^{\ell q}_{\alpha\beta}(s) = \frac{ e^2  Q_\ell Q_q}{s} + 
\frac{e^2 g_\alpha^\ell g_\beta^q}
{\sin^2\theta_{\rm w} \cos^2 \theta_{\rm w} } \;
\frac{1}{s - M_Z^2 } + \eta^{\ell q}_{\alpha \beta} \;,
\end{equation}
where $s,t,u$ are the usual Mandelstam variables.
In the above equations, $g_L^f = T_{3f} - Q_f \sin^2\theta_{\rm w}$,
$g_R^f = - Q_f \sin^2\theta_{\rm w}$, $Q_f$ is the electric charge of the
fermion $f$ in units of proton charge.  The SM amplitude can be recovered
by setting $\eta$'s to zero.

For a large class of new interactions the 
new physics contributions $\eta_{\alpha\beta}^{eq}$ vary slowly with $q^2$,
effectively being constant at energies accessible to present experiments, e.g.,
if the mass of the exchange quanta is much heavier than the energy scale of
the experiments.
In this case the $\eta_{\alpha\beta}^{ff'}$ correspond to constant four-fermion
contact interactions, and 
Eq.~(\ref{mab}) relates the sensitivity to new physics of all experiments
probing a given combination of external quarks and leptons, such as 
$ep\to eX$, $p\bar p\to e^+e^-X$, $e^+e^-\to \rm hadrons$ and atomic physics
parity violation experiments.
Based on the formula in Eq. (\ref{mab}) 
the amplitude squared for the deep-inelastic 
scattering at HERA can be obtained by a simple interchange of the Mandelstam
variables.

%
%

\section{Global Data}

The global data used in this analysis have been described in Ref. \cite{ours}.
Here we only describe those that have been updated since then.  We have used
the most recent CTEQ (v.5) parton distribution functions \cite{cteq5} wherever
they are needed.

\subsection{HERA data}

ZEUS \cite{zeus} and H1 \cite{H1} have published their results on the
NC deep-inelastic scattering (DIS)
at $e^+ p$ collision with $\sqrt{s} \approx 300$ GeV.  
The data sets of H1 and ZEUS are based on accumulated 
luminosities of 35.6 and 47.7 pb$^{-1}$, respectively.  
H1 \cite{H1} also published NC data for the most recent run of $e^- p$
collision at $\sqrt{s}\approx 320 $ GeV with an integrated luminosity of 
$16.4$ pb$^{-1}$.

We used the double differential cross section $d^2 \sigma/dx dQ^2$ given by 
the H1 \cite{H1} data and the single differential cross section 
$d\sigma/d Q^2$ given by ZEUS \cite{zeus} data in 
our fits.  At $e^+ p$ collision, the double differential cross section
for NC DIS, including the effect of $\eta$'s, is given by
\begin{eqnarray}
\frac{d^2\sigma}{dx dQ^2} (e^+p\to e^+X) &=&
 \frac{1}{16\pi}\; \Biggr \{
\sum_q f_q(x) \,\biggr [
     (1-y)^2 ( |M^{eq}_{LL}(t)|^2 + |M^{eq}_{RR}(t)|^2)  + 
      |M^{eq}_{LR}(t)|^2 + |M^{eq}_{RL}(t)|^2 \biggr ]
\nonumber  \\
&&+
\sum_{\bar q} f_{\bar q} (x) \, \biggr [
     |M^{eq}_{LL}(t)|^2 + |M^{eq}_{RR}(t)|^2  + 
    (1-y)^2 (|M^{eq}_{LR}(t)|^2 + |M^{eq}_{RL}(t)|^2) \biggr ]
 \; \Biggr \} \;, \label{nccc}
\end{eqnarray}
where $Q^2 = sxy$ is the square of the momentum-transfer 
and $f_{q/\bar q}(x)$ are parton distribution functions.  The reduced 
amplitudes $M_{\alpha\beta}^{eq}$ are given by Eq. (\ref{mab}).
The single differential cross section $d\sigma/d Q^2$ is obtained by 
integrating over $x$.
The corresponding formulas for $e^- p$ collision can be obtained from the 
above equation 
by interchanging $( LL \leftrightarrow LR, RR \leftrightarrow RL)$.

We normalize the tree-level SM cross section to the low $Q^2$ part 
of the data set by a scale factor $C$ (C is very close to 1.)  
The cross section $\sigma^{\rm th}$ used in the minimization procedure 
is then given by
\begin{equation}
\label{sign}
\sigma^{\rm th} = C \left( \sigma^{\rm SM} + \sigma^{\rm interf} +
\sigma^{\rm cont} \right )
\end{equation}
where $\sigma^{\rm interf}$ is the interference cross section between the
SM and the contact interactions and $\sigma^{\rm cont}$ is the cross
section due to contact interactions.

\subsection{Drell-yan Production}

Both CDF \cite{dy-cdf} and D\O\ \cite{dy-d0} 
measured the differential cross section 
$d\sigma/dM_{\ell\ell}$ for Drell-Yan production, 
where $M_{\ell\ell}$ is the invariant mass of the lepton pair.  
While CDF analyzed data from both electron and muon samples, D\O\ analyzed
only the electron sample. 

The differential cross section, including the contributions of contact
interactions, is given by
\begin{equation}
\frac{d^2\sigma}{dM_{\ell\ell} dy} = K \frac{M_{\ell\ell}^3}{72\pi s} \;
\sum_q f_q(x_1) f_{\bar q}(x_2)\; \left(
|M^{eq}_{LL}(\hat s)|^2 + |M^{eq}_{LR}(\hat s)|^2 + |M^{eq}_{RL}(\hat s)|^2 + 
|M^{eq}_{RR}(\hat s)|^2  \right ) \;,
\end{equation}
where $M_{\alpha\beta}^{eq}$ is given by Eq. (\ref{mab}), 
$\hat s=M_{\ell\ell}^2 $, $\sqrt{s}$ is the center-of-mass energy of the 
$p\bar p$ collision, $M_{\ell\ell}$ and $y$ are, respectively, 
the invariant mass and the rapidity of the lepton pair, and 
$x_{1,2} = \frac{M_{\ell\ell}}{\sqrt s} e^{\pm y}$, and $y$ is numerically
integrated.  The QCD $K$-factor is given by  $K=1+ \frac{\alpha_s(\hat s)}
{2\pi}\frac{4}{3}( 1+ \frac{4\pi^2}{3})$. 

We scale our tree-level SM cross section by normalizing to the $Z$-peak cross
section data.  The cross section used in the minimization procedure is then 
given similarly by Eq. (\ref{sign}).

\subsection{LEP}

The LEP Electroweak Working Group (LEPEW) 
combined the data on $q \bar q$ production
from the four LEP collaborations \cite{lepew} for energies between 
130 and 202 GeV.  In our previous fits, we have data upto 183 GeV only.
In the LEPEW report, they also noted that the hadronic cross section, on
average, is about $2.5 \sigma$ above the SM prediction.  In fact, we see this
effect in our fits.

In the report, both the experimental cross sections and predictions from the
next-leading-order (NLO) cross sections are given \cite{lepew}.  
Since the NLO  calculation for contact interactions is not 
available, we do the calculation
by first normalizing our tree-level results to the NLO cross sections
given in the report and then multiplying this scale factor to 
the new cross sections that include the SM and the contact interactions.

At leading order in the electroweak interactions, the total hadronic cross
section for $e^+e^-\to q\bar q$, summed over all flavors $q=u,d,s,c,b$, is
given by
\begin{equation}
\sigma_{\rm had} = K\, \sum_q \frac{s}{16\pi} \biggr[
|M_{LL}^{eq}(s)|^2 + |M_{RR}^{eq}(s)|^2 + |M_{LR}^{eq}(s)|^2 +
|M_{RL}^{eq}(s)|^2 \biggr] \,,
\label{sig_had}
\end{equation}
where $K=1+\alpha_s/\pi+1.409(\alpha_s/\pi)^2-12.77(\alpha_s/\pi)^3$
is the QCD $K$ factor.

We found that some of the fits are dominated by these $e^+ e^- \to
q \bar q$ hadronic cross sections.  If data at even higher energies $>202$ 
GeV are available, the limits will increase.
In our fits, we assumed a more conservative scenario that contact 
interactions only appear in $eu$ and $ed$ channels.  Have we assumed the
universalities of $eu=ec$ and $ed=es=eb$, the limits obtained would have been
significantly higher.

\subsection{Atomic Parity Violation}

The APV is measured in terms of weak charge $Q_W$.  The updated 
experimental value with an improved atomic calculation \cite{apv,atom} 
is about $1.0 \sigma$ larger than the SM prediction \cite{langacker}, namely,
$\Delta Q_W \equiv Q_W({\rm Cs}) - Q_W^{\rm SM}({\rm Cs}) 
= 0.44 \pm 0.44$.
The contribution to $\Delta Q_W$ from the contact parameters 
is given by \cite{bc,ours}
\begin{equation}
\label{QW}
\Delta Q_W = ( -11.4\; {\rm TeV}^{2} ) \left[
-\eta_{LL}^{eu} + \eta_{RR}^{eu} - \eta_{LR}^{eu} + \eta_{RL}^{eu} \right ]
+ 
( -12.8\; {\rm TeV}^{2} ) \left[
-\eta_{LL}^{ed} + \eta_{RR}^{ed} - \eta_{LR}^{ed} + \eta_{RL}^{ed} \right ]
\;.
\end{equation}
Note that the $\eta$'s come in special combinations.   If for some specific
combinations: e.g., vector-vector ($VV$): $\eta_{VV}=
\eta_{LL}=\eta_{LR}=\eta_{RL}= \eta_{RR}$ and 
axial-vector-axial-vector ($AA$):
$\eta_{AA}=\eta_{LL}=-\eta_{LR}=-\eta_{RL}=\eta_{RR}$, the contributions to
$\Delta Q_W$ are zero.  

There are also electron-nucleon scattering data, which have not been updated
since our previous fits.  The contributions to the asymmetries that were
measured in these experiments are automatically zero for similar combinations
of $\eta$'s.

\subsection{Charged-current (CC) Universality}

The difference 
$\eta^{ed}_{LL}-\eta^{eu}_{LL}=\eta_2/2$ measures the exchange of isospin
triplet quanta between left-handed leptons and quarks, as indicated by the 
presence of the $SU(2)$ generators $T^a=\sigma^a/2$ in the $\eta_2$ term. 
This term also provides an $e\nu ud$ contact term in CC processes. 
Such contributions, however, are severely restricted by lepton-hadron 
universality of weak charged currents \cite{altCC} within the experimental
verification of unitarity of the CKM matrix. The experimental 
values~\cite{pdg00}

\begin{equation}
|V_{ud}^{\rm exp}|=0.9735\pm 0.0008\;,\; 
|V_{us}^{\rm exp}|=0.2196\pm 0.0023\;,\;
|V_{ub}^{\rm exp}|=0.0036\pm 0.0010\;,
\end{equation}
lead to the constraint 
\begin{equation}
\left(|V^{\rm SM}_{ud}|^2+|V^{\rm SM}_{us}|^2+|V^{\rm SM}_{ub}|^2\right)
\left(1-{\eta_2\over 4\sqrt{2}G_F}\right)^2 = 0.9959\pm 0.0019\; ,
\end{equation}
when flavor universality of the contact interaction is assumed.
As a result $\eta_2$ must be small, though not necessarily negligible,
\begin{equation}\label{eq:eta2CC}
\eta_2 \equiv 2( \eta^{ed}_{LL} - \eta^{eu}_{LL} ) 
= (0.135\pm 0.063)\;{\rm TeV}^{-2}\;.
\end{equation}

Other data we used in our fits include 
low-energy electron-nucleon scattering experiments \cite{eN} and
neutrino-nucleon scattering experiments \cite{ccfr}.
When considering constraints from neutrino-nucleon
scattering experiments, we invoke the $SU(2)$ relations and $e$--$\mu$ 
universality in order to restrict 
the number of free parameters. In addition to using the relations of 
Eqs.~(\ref{su2releR}) and (\ref{su2relnu}), we will also impose the CC 
constraint on $\eta_2$ when neutrino data are included in the fits.

\section{Fits and Limits}

The fits of contact parameters are obtained by minimizing the $\chi^2$ of
the data sets.  In order to see how each data set affects the fit, we
first show the fits with each data set added one at a time, as 
shown in Table \ref{table1}.
We observe the following: (i) the SM model fits the data  well with
$\chi^2_{\rm SM}$/d.o.f. $\alt 1$ for all five columns in Table \ref{table1}.
(ii) The contact interaction fits the data slightly better than the SM.  In 
the last column of Table \ref{table1}, the $\chi^2_{\rm SM}$/d.o.f.$=0.975$
while $\chi^2_{\rm cont}$/d.o.f. $= 0.936$, where $\chi^2_{\rm SM}$ 
and $\chi^2_{\rm cont}$ are the chi-square for the SM and 
contact interactions, respectively.
(iii) The $\chi^2_{\rm cont}$ for APV in all cases are zero, which means that
the minimization procedure prefers the APV data to be satisfied.  In other
words, other choices of $\eta$'s would give a too large  $\chi^2$ if APV data
is violated to a large extent.

In view of these, we conclude that the global data do not show any sign 
of contact interactions.  Thus, we can derive 95\% C.L. limits on the 
compositeness scale, below which the contact interaction is ruled out.
The 95\% C.L. one-sided limits $\eta^{95}_\pm$ are defined, respectively, as
\begin{equation}
0.95 = \frac{\int_0^{\eta^{95}_+} \;d \eta \; P(\eta) }
            {\int_0^\infty       \; d \eta \; P(\eta) } \qquad  {\rm and}
\qquad 
0.95 = \frac{\int^0_{\eta^{95}_-} \;d \eta \; P(\eta) }
            {\int^0_{-\infty}   \; d \eta \; P(\eta) } \;
\end{equation}
where $P(\eta)$ is the fit likelihood given by 
$P(\eta)= \exp ( - (\chi^2(\eta) - \chi^2_{\rm min})/2 )$.   
The 95\% C.L. limits on
$\Lambda_\pm = \sqrt{ \pm \frac{4\pi}{\eta^{95}_\pm} }$.
The limits on $\Lambda_{\pm}$ are summarized
 in Tables \ref{table2}--\ref{table4}.
In Table \ref{table2}, for each chirality coupling 
considered the others are put to
zero. The limits on $\Lambda$ obtained range from 10--26 TeV, which improve
significantly from each individual experiment.
We also calculate the limits on the compositeness scale when some symmetries
on contact terms are considered, as shown in Table \ref{table3}:
$VV$ stands for vector-vector:
$\eta_{LL}=\eta_{LR}=\eta_{RL}=\eta_{RR}=\eta_{VV}$, while
$AA$ stands for axial-vector-axial-vector:
$\eta_{LL}=-\eta_{LR}=-\eta_{RL}=\eta_{RR}=\eta_{AA}$.
These limits, in general, are 
not as strong as those in the previous table because the additional symmetry
automatically satisfies the parity violation experiments: APV and $e$-N.

Finally, we show the limits that can be obtained from each set of data by
looking at the results of $LL$ and $VV$ cases.  The 
former is constrained severely by the APV and CC data, while the latter is free
from the APV data.  For the $LL$ case the most dominant constraint is the
CC universality, followed closely by the APV data (as indicated by the
error of the best fit values.)  The stringencies of the CC universality is
understood because the $LL$ interaction affects the $V-A$ structure.  The CC
universality will not constrain chirality combinations other than $LL$.  
On the other hand, parity-violating experiments will not be able to constrain
the $VV$ case, and 
the strongest constraint then comes from the LEP hadronic cross sections.

\section{Implications to Leptoquark models}

The interaction Lagrangians for the $F=0$ and $F=-2$ 
($F$ is the fermion number) scalar leptoquarks are \cite{buch}
\begin{eqnarray}
\label{9}
{\cal L}_{F=0} &=& \lambda_L \overline{\ell_L} u_R {\cal S}_{1/2}^L
+ \lambda_R^* \overline{q_L} e_R (i \tau_2 {\cal S}^{R*}_{1/2} )
+ \tilde{\lambda}_L \overline{\ell_L} d_R \tilde{{\cal S}}_{1/2}^L + h.c. \;,\\
\label{10}
{\cal L}_{F=-2} &=& g_L \overline{q_L^{(c)}} i \tau_2 \ell_L {\cal S}_0^L
+ g_R \overline{u_R^{(c)}} e_R {\cal S}_0^R
+ \tilde{g}_R \overline{d_R^{(c)}} e_R \tilde{{\cal S}}_0^R
+ g_{3L}\overline{q_L^{(c)}} i \tau_2 \vec{\tau} \ell_L \cdot \vec{\cal S}_1^L
+ h.c.
\end{eqnarray}
where $q_L,\ell_L$ denote the left-handed quark and lepton doublets, 
$u_R,d_R,e_R$ denote the right-handed up-type quark, down-type quark, and 
lepton singlet, and $q_L^{(c)}, u_R^{(c)}, d_R^{(c)}$ denote the 
charge-conjugated fields.
The subscript on leptoquark fields denotes the weak-isospin of the leptoquark,
while the superscript ($L,R$) denotes the handedness of the lepton that
the leptoquark couples to.  The color indices of the quarks and leptoquarks
are suppressed.  Note that the above Lagrangians have the SU(2)$_L$ symmetry
and thus obey the SU(2) relations in Eqs. (\ref{su2releR}) and 
(\ref{su2relnu}).

It is convenient to express the effects of leptoquarks in terms of the
contact interaction coefficients $\eta$'s.  This is made possible when
the mass of the leptoquark is much larger than the momentum transfer in the
process.  We classify the effects as follows. \\
(i) ${\cal S}^{L,R}_{1/2}$: 
\begin{eqnarray}
\eta^{eu}_{LR} &=& - \frac{|\lambda_L|^2}{2 M^2_{{\cal S}_{1/2}}} 
=\eta^{\nu u}_{LR} \;, \nonumber \\
\eta^{eu}_{RL} &=& - \frac{|\lambda_R|^2}{2 M^2_{{\cal S}_{1/2}}} 
=\eta^{e d}_{RL} \;.
\end{eqnarray}
(ii) $\tilde{\cal S}^L_{1/2}$:
\begin{equation}
\eta^{ed}_{LR} = - \frac{|\tilde{\lambda}_L|^2}
{2 M^2_{\tilde{\cal S}_{1/2}}} 
=\eta^{\nu d}_{LR} \;.
\end{equation}
(iii) ${\cal S}^{L,R}_{0}$: 
\begin{eqnarray}
\eta^{eu}_{LL} &=& \frac{|g_L|^2}{2 M^2_{{\cal S}_{0}}} 
=\eta^{\nu d}_{LL} \;, \nonumber \\
\eta^{eu}_{RR} &=&  \frac{|g_R|^2}{2 M^2_{{\cal S}_{0}}} \;.
\end{eqnarray}
(iv) $\tilde{\cal S}^R_{0}$:
\begin{equation}
\eta^{ed}_{RR} =  \frac{|\tilde{g}_R|^2}
{2 M^2_{\tilde{\cal S}_{0}}} \;.
\end{equation}
(v) ${\vec S}^{L}_{0}$: 
\begin{eqnarray}
\eta^{eu}_{LL} &=& \frac{|g_{3L}|^2}{2 M^2_{{\vec S}_{1}}} 
=\eta^{\nu d}_{LL} \;, \nonumber \\
\eta^{ed}_{LL} &=& \eta_{LL}^{\nu u} = 2 \eta_{LL}^{eu} \;.
\end{eqnarray}

Once we expressed the effects in terms of $\eta$'s, we can directly 
analyze the combinations of $\eta$'s in the global fit.  The resulting limits
are given in terms of $\lambda^2/2M^2_{\rm LQ}$.  Conventionally, the 
coupling constants $\lambda_{L,R}$ or $g_{L,R,3L}$ are assumed the 
electromagnetic strength, i.e., $\lambda_{L,R}=e=g_{L,R,3L}$ and thus we 
can obtain the lower limits on $M_{\rm LQ}$.  These results are summarized in
Table \ref{table-lq}.  
Roughly, the leptoquark masses are required to be larger than 1 TeV, in 
order to satisfy all the constraints (except that ${\cal S}^{L,R}_0$ has to be 
heavier than 1.7 TeV and ${\cal S}^{L,R}_{1/2}$ can be as light as
0.67 TeV)  when the coupling constants are assumed an 
electromagnetic strength $e$.  The strongest constraint comes from APV and
CC universality, the latter of which constrains the $LL$ chirality severely.

This is an interesting result in view of the
recent measurement of muon anomalous magnetic moment \cite{E821} with 
respect to a couple of leptoquark solutions \cite{LQ}.  The most favorable
leptoquark solution to the muon anomaly is ${\cal S}^{L,R}_{1/2}$ that has
both left- and right-handed couplings with the allowed mass range in
$0.8\;{\rm TeV} < M_{{\cal S}_{1/2}} < 2.2\; {\rm TeV}$.  This solution is
in total consistency with the global NC constraint (as shown in (i) of
Table \ref{table-lq}.)

\section{Implications to $Z'$  models}

We can write down the Lagrangian of a generic $Z'$ model coupling to 
fermions as
\begin{equation}
{\cal L} = - g_E  \sum_f \;  \bar f \gamma^\mu \, \left(
\epsilon_L(f) \, P_L  +  \epsilon_R(f) P_R \, \right ) \, f \; Z'_\mu \;,
\end{equation}
where $P_{L,R}= (1\mp \gamma^5)/2$, 
$g_E= \sqrt{5\lambda_g/3}\, e/\cos\theta_{\rm w}$ and $\lambda_g$ is typically
in the range $2/3-1$ and for grand-unified theories breaking directly into
SU(3)$\times$SU(2)$\times$U(1)$\times$U(1)' $\lambda_g=1$, 
and $\epsilon_{L,R}(f)$ are the left- and right-handed chiral couplings to the
$Z'$.
Here in this simple analysis, we assume that the $Z'$ does not mix with 
the SM $Z$ boson such that the $Z'$ is not constrained by the electroweak
precision data \cite{cho+hagi}.

The contribution of the $Z'$ to the reduced amplitudes
is given by
\begin{equation}
M_{\alpha\beta}^{eq} = \left.  M_{\alpha\beta}^{eq} \right |_{\rm SM} +
 \frac{g_E^2} {q^2 - M^2_{Z'}} \; \epsilon_\alpha(e) \, \epsilon_\beta(q)
\;.
\end{equation}
In other words, if $M^2_{Z'} \gg q^2$ the effects of $Z'$ can be expressed
in terms of the contact interaction parameters $\eta$'s as
\begin{equation}
\label{assume}
\eta_{\alpha\beta}^{eq} = - \frac{g_E^2}{M^2_{Z'}} \;
\epsilon_\alpha(e) \, \epsilon_\beta(q) \;.
\end{equation}

Once we expressed the effects of $Z'$ in terms of contact interaction
parameters, we can easily analyze the $Z'$ models in our global fit. 
We shall analyze the following $Z'$ models  ($\lambda_g=1$) \cite{paul}:\\
(i) Sequential $Z$ model:
\begin{equation}
\epsilon_{L,R}(f)  = T_{3f} - Q_f \sin^2 \theta_{\rm w} \;.
\end{equation}
(ii) Left-right $Z_{LR}$ model:
\begin{equation}
\epsilon_L(f) = \sqrt{\frac{3}{5}} \, \left( \frac{-1}{2\alpha} \right )
\,(B-L)_f \,, \;\;
\epsilon_R(f) = \sqrt{\frac{3}{5}} \left[  \alpha T_{3R}^f - 
\frac{1}{2\alpha} \, (B-L)_f \right ] \,,  \;\;
\alpha = \sqrt{\frac{1-2 \sin^2\theta_{\rm w} }{ \sin^2 \theta_{\rm w}}} \,. 
\end{equation}
(iii) $Z_\chi$ model:
\begin{equation}
\epsilon_L(u) = -\epsilon_R(u) = \epsilon_L(d) =
\frac{\epsilon_R(d)}{3} = -\frac{\epsilon_L(e)}{3} = - \epsilon_R(e) =
-\frac{\epsilon_L(\nu)}{3} = - \frac{1}{2\sqrt{10}} \;.
\end{equation}
(iv) $Z_\psi$ model:
\begin{equation}
\epsilon_L(u) = -\epsilon_R(u) = \epsilon_L(d) =
-\epsilon_R(d) = \epsilon_L(e) = - \epsilon_R(e) = \epsilon_L(\nu) = 
\frac{1}{\sqrt{24}} \;.
\end{equation}
(v) $Z_\eta$ model:
\begin{equation}
\epsilon_L(u) = -\epsilon_R(u) = \epsilon_L(d) =
2 \epsilon_R(d) = -2 \epsilon_L(e) = - \epsilon_R(e) = - 2\epsilon_L(\nu) = 
- \frac{2}{2\sqrt{15}} \;. 
\end{equation}
Note that the $Z_\psi$ gives an axial-vector-axial-vector interaction, which
evades strong constraints of APV and CC universality. In fact, $Z_\chi$ and
$Z_\eta$ are also not constrained by CC universality.  The resulting best
estimates of $g_E^2/M^2_{Z'}$ for each model are shown in Table \ref{table-z},
with the corresponding lower limits on $Z'$ masses. 

The results shown in the Table are not satisfactory.  First, the lower mass
limits on $Z_\psi$ and $Z_\eta$ are rather low, 0.16 and 0.43 TeV respectively.
We have verified that $Z_\psi$ gives only AA-type interactions and the 
result of $Z_\psi$ is consistent with $\eta^{eq}_{AA}$ of Table \ref{table3}.
Such low mass values invalidate the assumption of Eq. (\ref{assume}), 
which means that we cannot apply the simple contact interaction analysis 
to these $Z'$ models.  In this case, more sophisticated $q^2$-dependent 
analysis is necessary to get an accurate result, which is beyond the scope 
of the present paper
\footnote{The CDF and D\O\ collaborations \cite{Zprime}
have done such $q^2$-dependent analyses on Drell-yan production
to put limits on $Z'$ models.}.
Nevertheless, it should be reasonably applicable to sequential 
$Z$ model, $Z_{LR}$ and $Z_\chi$.

\section{Conclusions}

In conclusion, we have examined the NC $eeqq$ data and found that 
the data do not 
support the existence of $eeqq$ contact interactions with the compositeness
scale upto 6--26 TeV, depending on the chiralities.  We have also demonstrated
that the low-energy data (APV and CC universality) dominate the fit for the
$LL$ chirality.  In the case of parity-conserving contact interactions, the 
LEP hadronic cross section  dominates the fit. 

The above analysis has also been applied in a straight-forward fashion to 
other new physics such as leptoquark and $Z'$ models.
For leptoquark models we found the 95\% C.L. lower mass limits range from
0.67 to 1.7 TeV.  Especially, the leptoquark ${\cal S}^{L,R}_{1/2}$, which
couples to both left- and right-handed charged leptons, 
has a mass limit of 0.67
TeV, which is consistent with the best leptoquark solution to the muon 
anomalous magnetic moment anomaly.
For $Z'$ models we found that our analysis is applicable to the sequential $Z$,
$Z_{LR}$, and $Z_\chi$ models, with mass limits ranging from 0.68 to 1.5 TeV.
We found that our analysis is not applicable to $Z_\psi$ and $Z_\eta$ models.

I would like to thank Vernon Barger, Karou Hagiwara, and Dieter Zeppenfeld for
previous collaborations, which lead to the present work. 
This research was supported in part by the National Center for Theoretical
Science under a grant from the National Science Council of Taiwan R.O.C.



\newpage
\begin{table}[t]
\caption[]{\label{table1}
\small
The best estimate of the $\eta_{\alpha\beta}^{eq}$ parameters in units of
TeV$^{-2}$ when various
data sets are added successively.  In the last column when the $\nu$-N
data are included the $\eta_{L\beta}^{\nu q}$ are given in terms of
$\eta_{L\beta}^{eq}$ by Eq.~(\protect\ref{su2relnu}) and
we assume $\eta_{RL}^{eu}=\eta_{RL}^{ed}$ in the last column.
}
\medskip
\centering 
\begin{tabular}{|l|c|c|c|c|c|}
\hline
 & HERA only & HERA+APV & HERA+APV & HERA+APV   &  HERA+DY+APV \\
 &           & +eN      & +eN+DY   &+eN+DY+LEP & +eN+LEP+$\nu$N+CC \\
\hline
$\eta_{LL}^{eu}$ & $-2.10\err{1.74}{1.76}$
  & $-1.74\err{1.14}{1.04}$  & $0.09\err{0.39}{0.31}$
  & $0.07\err{0.37}{0.30}$  & $0.01\pm{0.20}$ \\
$\eta_{LR}^{eu}$ & $-2.49\err{1.49}{1.18}$ 
  & $-1.52\err{0.98}{1.12}$  & $-0.16\err{0.40}{0.38}$
 & $-0.15\err{0.40}{0.38}$  & $- 0.21\err{0.22}{0.23}$ \\
$\eta_{RL}^{eu}$ & $-1.53\err{1.53}{1.38}$ 
  & $-2.47\err{1.11}{0.80}$  & $-0.33\err{0.43}{0.39}$
  & $-0.32\err{0.42}{0.39}$  & $-0.22\pm{0.32}$ \\
$\eta_{RR}^{eu}$ & $-1.19\err{1.81}{1.83}$  
 & $-1.37\err{0.98}{1.09}$ & $-0.16\err{0.41}{0.36}$
 & $-0.19\err{0.38}{0.33}$ & $-0.17\err{0.36}{0.26}$ \\
$\eta_{LL}^{ed}$ & $-5.35\err{2.88}{2.30}$
 & $-6.39\err{1.08}{0.92}$  & $-0.21\err{0.88}{0.69}$
 & $-0.25\err{0.55}{0.52}$  & $0.08\pm{0.21}$ \\
$\eta_{LR}^{ed}$ & $-1.24\err{0.80}{0.79}$
 & $-1.48\err{0.77}{0.75}$  & $-0.93\pm {0.54}$
 & $-0.90\err{0.53}{0.51}$  & $-0.62\err{0.34}{0.32}$ \\
$\eta_{RL}^{ed}$ & $-3.62\err{1.97}{1.30}$ 
 & $-2.41\err{1.56}{1.28}$  & $-0.89\err{0.98}{0.76}$
 & $-0.78\err{0.91}{0.64}$  & $=\eta_{RL}^{eu}$ \\
$\eta_{RR}^{ed}$ & $-6.01\err{3.18}{2.63}$ 
 & $-4.97\err{1.63}{1.51}$  & $0.08\err{0.92}{0.90}$
 & $-0.02\err{0.47}{0.54}$  & $-0.18\err{0.41}{0.48}$ \\
\hline
\hline
HERA    & 230.4& 231.3  & 250.9 & 250.8  & 251.3  \\
APV     &      & 0.0    & 0.0   &  0.0 & 0.0  \\
eN      &      & 0.6    & 0.7   &  0.7  & 1.4  \\
DY      &      &        & 51.0  &  51.2  & 51.0 \\
LEP     &      &        &       &  4.2  & 4.2 \\
$\nu$-N  &      &        &       &        &  0.1   \\
\hline
\hline
Total $\chi^2$ ($\chi^2_{\rm cont}$)
    & 230.4 & 232.0   & 302.6 &  306.9 & 307.9  \\
\hline
SM $\chi^2$ ($\chi^2_{\rm SM}$) & 257.4 &  260.3 & 311.1 & 321.7 & 327.7 \\
\hline
SM d.o.f.   & 276    &  281     & 323    & 333     & 336 \\
\hline
\end{tabular}
\end{table}

\begin{table}[th]
\caption[]{\label{table2}
\small
The best estimate on $\eta_{\alpha\beta}^{eq}$ and the 95\% C.L. limits on the
compositeness scale $\Lambda_{\alpha\beta}^{eq}$,
where $\eta_{\alpha\beta}^{eq}=4\pi\epsilon/(
\Lambda_{\alpha\beta\epsilon}^{eq})^2$.
When one of the $\eta$'s is considered the others are set to zero.
SU(2) relations are assumed and $\nu$N and CC data are included.
}
\medskip
\centering
\begin{tabular}{|cc|cc|}
\hline
& & \multicolumn{2}{c|}{95\% C.L. Limits} \\
Chirality  ($q$) &  Best estimate (TeV$^{-2}$)  &  
$\Lambda_+$ (TeV) & $\Lambda_-$ (TeV) \\
\hline
\hline
LL($u$)  & $-0.044 \pm 0.022$ &  23.3  & 12.5 \\
LR($u$)  & $0.027  \pm 0.038$ &  11.6  & 14.8 \\
RL($u$)  & $-0.023\pm 0.018$ &  23.1 & 15.2 \\
RR($u$)  & $-0.073 \pm 0.037$ &  17.9 & 9.7 \\
LL($d$)  & $0.065 \pm 0.022$ &  11.1  & 26.4 \\
LR($d$)  & $0.031 \pm 0.034$ &  11.7  & 15.9\\
RR($d$)  & $-0.021\pm 0.034$ &  15.2 & 12.3 \\
\hline
\end{tabular}
\end{table}

\begin{table}[th]
\caption[]{\label{table3}
\small
The best estimate on $\eta^{eq}$ for $VV,AA$, and
the corresponding 95\% C.L. limits on the
compositeness scale $\Lambda$, where $\eta=4\pi\epsilon/
(\Lambda_{\epsilon})^2$.
When one of the $\eta$'s is considered the others are set to zero.
When we consider contact terms for just the $u$ or $d$, we cannot apply
SU(2) relations and so we do not include the $\nu$N and CC data.  On the 
other hand, when both $u$ and $d$ contact terms are considered, we can apply
the SU(2) relations and thus include the $\nu$N and CC data.
}
\medskip
\centering
\begin{tabular}{|cc|cc|}
& & \multicolumn{2}{c|}{95\% C.L. Limits} \\
Chirality combinations  &  Best estimate (TeV$^{-2}$) &
$\Lambda_+$ (TeV)  & $\Lambda_-$ (TeV) \\
\hline
\hline
$\eta_{VV}^{eu}$   & $-0.12\err{0.037}{0.036}$ &  20.0 & 8.4 \\
$\eta_{VV}^{ed}$   & $0.19\err{0.068}{0.072}$ &  6.6 & 12.2 \\
\hline
$\eta_{AA}^{eu}$  & $-0.15\err{0.054}{0.052}$ & 15.0  &  7.3 \\
$\eta_{AA}^{ed}$  & $0.18 \err{0.055}{0.058}$   & 6.9  &  15.1 \\
\hline
$\eta_{VV}^{eu}=\eta_{VV}^{ed}$   
                & $-0.18\err{0.051}{0.048}$  &  15.7  & 7.0 \\
$\eta_{AA}^{eu}=\eta_{AA}^{ed}$  
                & $-0.20\err{0.15}{0.099}$  & 5.7   & 6.1 \\
\end{tabular}                                               
\end{table}

\begin{table}[th]
\caption[]{\label{table4}
\small
The best estimate on $\eta^{eq}_{LL}$ and $\eta^{eq}_{VV}$ for
each set of data as shown.  The corresponding 95\% C.L. lower limits on
the compositeness scale $\Lambda$ are also shown.
}
\medskip
\centering
\begin{tabular}{|c||cc|cc|cc|cc|}
 & \multicolumn{2}{c|}{HERA NC} & \multicolumn{2}{c|}{Drell-yan}  & 
 \multicolumn{2}{c|}{LEP $\sigma_{\rm had}$} & 
 \multicolumn{2}{c|}{APV+eN+$\nu$N+CC} \\
\hline
& \underline{$\eta$ (TeV$^{-2}$)} & 
\underline{$\Lambda_{+} / \Lambda_{-}$ (TeV)} 
& \underline{$\eta$} & \underline{$\Lambda_{+} / \Lambda_{-}$} 
& \underline{$\eta$} & \underline{$\Lambda_{+} / \Lambda_{-}$} 
& \underline{$\eta$} & \underline{$\Lambda_{+} / \Lambda_{-}$}  \\
$\eta_{LL}^{eu}$  & $-1.18 \err{0.53}{0.56}$ & $5.3/2.4$  & 
                    $-0.19 \err{0.24}{0.21}$ & $5.1/4.9$  & 
                    $-0.22 \err{0.086}{0.084}$ & $12.3/5.9$  & 
                    $-0.028 \pm{0.023}$ & $20.6/13.7$    \\
$\eta_{LL}^{ed}$  & $1.53 \err{1.59}{1.35}$ & $1.6/2.9$  & 
                    $0.88 \err{0.58}{0.73}$ & $2.7/2.7$  & 
                    $0.26 \err{0.095}{0.098}$ & $5.6/11.4$  & 
                    $0.054 \pm{0.022}$ & $11.7/24.4$    \\
$\eta_{LL}^{eu}=\eta_{LL}^{ed}$
                  & $-4.75 \err{1.56}{1.13}$ & $4.7/1.4$  & 
                    $-0.19 \err{0.32}{0.24}$ & $3.4/4.8$  & 
                    $-0.69 \err{0.19}{0.16}$ & $3.0/3.7$  & 
                    $0.017 \pm{0.018}$ & $16.0/22.0$    \\
\hline
\hline
$\eta_{VV}^{eu}$   & $-0.30\pm 0.13$ & $10.3/4.9$ &
                     $-0.054 \err{0.12}{0.11}$ & $6.7/7.4$  & 
                     $-0.11\err{0.042}{0.041}$ & $17.5/8.4$ & 
                      - & - \\
$\eta_{VV}^{ed}$  &  $-0.47\err{0.50}{0.48}$ & $4.1/3.2$ & 
                     $0.34 \err{0.41}{1.27}$ & $3.7/3.0$ & 
                     $0.20\err{0.068}{0.072}$ & $6.5/2.4$ & - & - \\
$\eta_{VV}^{eu}=\eta_{VV}^{ed}$  & $-0.38\err{0.14}{0.15}$ & $10.5/4.5$ & 
                                   $-0.060\err{0.15}{0.11}$& $5.0/7.2$ & 
                                   $-0.19\err{0.068}{0.061}$ & $3.3/6.6$ & 
                                   $-0.053\err{0.23}{0.27}$ & $5.8/1.9$
\end{tabular}
\end{table}

\begin{table}[th]
\caption[]{\label{table-lq}
\small
The best estimate on the leptoquark parameter $\eta=\lambda^2/2M^2_{\rm LQ}$, 
as well as the 95\% C.L. upper limits on $\eta$.
The corresponding 95\% C.L. lower limits on the leptoquark mass are also 
shown, with the coupling constants $\lambda_{L,R}=g_{L,R,3L}=e$ assumed.
The SU(2) relations are applied and $\nu$N and CC data are included in the
global analysis.
}
\medskip
\centering
\begin{tabular}{|l||c|cc|c|}
& Best Estimate  (TeV$^{-2}$) 
& \multicolumn{2}{l|}{95\% C.L. upper limits on $\eta$}
& 95\% C.L. lower limit \\
& &  \underline{$\eta^{95}_+$} & \underline{$\eta^{95}_-$} 
  & on $M_{\rm LQ}$ (TeV) \\
\hline
&&&& \\
(i) ${\cal S}^{L,R}_{1/2}$:  &  & & &  \\
$\eta^{eu}_{LR} = \eta^{eu}_{RL}$ & $-0.055 \pm 0.033$ & $0.038$ & $-0.11$ &
         0.67 \\ 
\hline
&&&& \\
(ii) $\tilde{\cal S}^{L}_{1/2}$:  &  & & &  \\
$\eta^{ed}_{LR} $ & $0.031 \pm 0.034$ & $0.091$ & $-0.050$ &
         1.0 \\ 
\hline
&&&& \\
(iii) ${\cal S}^{L,R}_{0}$:  &  & & &  \\
$\eta^{eu}_{LL} = \eta^{eu}_{RR}$ & $-0.087 \pm 0.024$ & $0.018$ & $-0.13$ &
         1.7 \\ 
\hline
&&&& \\
(iv) $\tilde{\cal S}^{R}_{0}$:  &  & & &  \\
$\eta^{ed}_{RR} $ & $-0.021 \pm 0.034$ & $0.055$ & $-0.083$ &
         0.94 \\ 
\hline
&&&& \\
(v) ${\vec S}^{L}_{0}$:  &  & & &  \\
$\eta^{eu}_{LL} = \eta^{ed}_{LL}/2$ & $0.022 \pm 0.011$ & $0.040$ & $-0.012$ &
         1.1 
\end{tabular}
\end{table}

\begin{table}[th]
\caption[]{\label{table-z}
\small
The best estimate on the $Z'$ model parameter $\eta=g_E^2/M^2_{Z'}$, 
as well as the 95\% C.L. upper limits on $\eta$.
The corresponding 95\% C.L. lower limits on the $Z'$ mass are also 
shown.
The SU(2) relations are applied and $\nu$N and CC data are included in the
global analysis.
$^*$ in the table denotes that these values invalidate the assumption of 
Eq. (\ref{assume}) and thus not valid.
}
\medskip
\centering
\begin{tabular}{|l||c|cc|c|}
& Best Estimate  (TeV$^{-2}$) 
& \multicolumn{2}{l|}{95\% C.L. upper limits on $\eta$}
& 95\% C.L. lower limit \\
& &  \underline{$\eta^{95}_+$} & \underline{$\eta^{95}_-$} 
  & on $M_{Z'}$ (TeV) \\
\hline
&&&& \\
(i) Sequential $Z$  & $-0.41 \pm 0.12$ & $0.095$ & $-0.62$ & 1.5 \\ 
\hline
&&&& \\
(ii) $Z_{LR}$ & $0.001 \pm 0.15$ & $0.29$ & $-0.28$ & 0.86 \\ 
\hline
&&&& \\
(iii) $Z_\chi$ & $0.17 \pm 0.17$ & $0.46$ & $-0.24$ & 0.68 \\ 
\hline
&&&& \\
(iv) $Z_\psi$ & $4.91 \err{2.38}{3.65}$ & $8.18$ & $-9.32$ & 0.16$^*$ \\ 
\hline
&&&& \\
(v) $Z_\eta$ & $-0.21 \err{0.66}{0.67}$ & $1.16$ & $-1.46$ & 0.43$^*$  
\end{tabular}
\end{table}

\end{document}